\magnification=\magstep1
\catcode`"=11
\let\quote="
\catcode`"=12
\chardef\foo="22
\global\newcount\refno \global\refno=1
\newwrite\rfile
\newlinechar=`\^^J
\def\@ref#1#2{\the\refno\n@ref#1{#2}}
\def\h@ref#1#2#3{\href{#3}{\the\refno}\n@ref#1{#2}}
\def\n@ref#1#2{\xdef#1{\the\refno}%
\ifnum\refno=1\immediate\openout\rfile=\jobname.refs\fi%
\immediate\write\rfile{\noexpand\item{[\noexpand#1]\ }#2.}%
\global\advance\refno by1}
\def\nref{\n@ref} 
\def\ref{\@ref}   
\def\hrref{\h@ref}
\def\lref#1#2{\the\refno\xdef#1{\the\refno}%
\ifnum\refno=1\immediate\openout\rfile=\jobname.refs\fi%
\immediate\write\rfile{\noexpand\item{[\noexpand#1]\ }#2\semi}%
\global\advance\refno by1}
\def\cref#1{\immediate\write\rfile{#1\semi}}

\def\preref#1#2{\gdef#1{\@ref#1{#2}}}

\def\semi{;\hfil\noexpand\break}

\def\listrefs{\vskip 20pt\immediate\closeout\rfile
\centerline{{\bf References}}\bigskip\frenchspacing%
\input \jobname.refs\vfill\eject\nonfrenchspacing}


\def\PRL{Phys.\ Rev.\ Lett.\ }

\global\newcount\meqno \global\meqno=1
\def\eqn#1{\warnIfChanged#1{\the\meqno}%
  \eqno(\the\meqno)\xdef#1{\the\meqno}%
     \global\advance\meqno by1
\eqlabeL#1%
\ifWritingAuxFile\immediate\write\auxfile{\noexpand\xdef\noexpand#1{#1}}\fi%
}
\def\defeqn#1{\warnIfChanged#1{\the\meqno}%
  \xdef#1{\the\meqno}%
     \global\advance\meqno by1
\eqlabeL#1%
\ifWritingAuxFile\immediate\write\auxfile{\noexpand\xdef\noexpand#1{#1}}\fi%
}
\def\anoneqn{\eqno(\the\meqno)%
     \global\advance\meqno by1
}

\baselineskip 15pt
\def\sect#1{\vskip 20pt\leftline{\bf #1}\vskip 10pt}
\def\mref#1#2{${}^{\the\refno)}$\n@ref#1{#2}}
\def\cite#1{${}^{#1)}$}

\hsize=17truecm
\vsize=24.8truecm
\hoffset=-5truemm
\voffset=-2truemm
\nopagenumbers

\topskip 2truecm
\rightline{hep-ph/yyymmdd}
\rightline{Saclay/SPhT--T96/051}
\vskip 18truemm
\centerline{{\bf OVERVIEW OF PERTURBATIVE QCD}\footnote{${}^\dagger$}{%
\sevenrm Presented at the XXXI${}^{\rm st}$ Rencontres de Moriond,
March 23--30, 1996, Les Arcs, France}}

\vskip 30truept
\centerline{David A. Kosower}
\centerline{Service de Physique Th\'eorique,
{CEA--Saclay}}
\centerline{F-91191 Gif-sur-Yvette cedex, France}

\vskip 4truecm
\centerline{\bf Abstract}
\vskip 20pt

I review theoretical techniques and current issues in perturbative
QCD, primarily as applied to jet physics at colliders.
\vfil

\eject
\topskip 0mm
\leftline{\bf Introduction}
\vskip 10pt

In the not-too-distant past, a talk such as this would have discussed
``tests of QCD.''  Perturbative QCD passed those tests.  What, then, is
its future in collider physics?
As we all know, the ultimate fate of yesterday's hot physics,
at least in particle physics, is to become tomorrow's background to newer,
and presumably hotter, physics.  Is this the destiny of perturbative QCD?

To a certain extent, this has already happened.  In the analysis leading
up to the unveiling of the top quark, the single-lepton channel --- where
one of the top-antitop pair decays hadronically, while the other's
daughter $W$ boson decays leptonically --- played a crucial role.  The analysis
in this channel requires a careful study of the QCD backgrounds arising
from jet production in association with a lone $W$.

But rather than merely fighting the QCD backgrounds, we may hope to use
jet physics as a tool in searching for new physics.  Refining
our theoretical and experimental techqniues with this idea in mind will
be one of the important challenges in the coming decade leading up to
the commissioning of the LHC.  In addition, we may hope to use jet
physics to extract information about nonperturbative quantities, such
as the parton distribution functions of the nucleon, that remain beyond
the present reach of lattice calculations.

Within the context of hadron colliders, experimenters are pursuing
studies of a wide variety of jet-associated final states: pure jet
production, production of photons or electroweak vector bosons in
association with jets, inclusive production of heavy quarks,
production of heavy-quark mesons, and production of quarkonia.  Different
distributions have applications to detailed studies of
standard-model observables, such as the mass of the top quark or
the mass of the $W$ boson.  They also have applications
 to measurements of parton distributions
of the nucleon, especially the gluon distribution, as well as to searches
for higher-dimension operators (such as might arise from compositeness or
the presence of heavy colored particles in a shorter-distance theory),
and to searches for speculative extensions of the standard model.

Jet studies at $ep$ machines can also be useful sources of information
about parton distributions, while those at high-statistics $e^+e^-$
machines will offer, upon completion of the next generation of theoretical
calculations, a precise measurement of the strong coupling $\alpha_s$.

Refining jet physics as a tool for doing physics at colliders will
require a great deal of theoretical work: in one-loop
matrix element calculations; in writing general-purpose, fully-differential
numerical programs for a larger number of processes; in setting up
a framework for giving honest estimates of errors
in predictions due to uncertainties in the extraction of parton distributions;
in performing yet-higher order calculations in order to give honest
theoretical error estimates.  It will also require experimenters to
focus on measuring and analyzing those observables which can be
predicted most reliably in perturbation theory.

\vfill\eject
\sect{Next-to-Leading Order Calculations}

Leading-order calculations
perturbative QCD rely only on tree-level matrix elements.  These calculations
provide a basic description of cross sections and distributions, but
are sensitive to potentially large, but uncalculated logarithms.  In
addition, other aspects of jets, such as their internal structure,
cannot be calculated at all in a leading-order program.

The presence of ultraviolet logarithms is reflected in the
residual renormalization-scale dependence of a perturbative prediction.
Truncating  a perturbative expansion at finite order introduces
a spurious dependence of a physical observable on the renormalization
scale $\mu_R$.  Next-to-leading order (NLO)
calculations can, and in practice usually
do, reduce this dependence significantly compared with leading-order ones.
At leading order, the only dependence on $\mu_R$ comes from the resummation
of logarithms in the running coupling $\alpha_s(\mu_R)$, and the scale choice
is arbitrary.  At NLO, in contrast, the virtual corrections
to the matrix element introduce a separate dependence on $\mu_R$.

As an aside, I would like caution against the common practice of assigning
a theoretical ``error'' by varying the scale up and down by a certain
factor (typically two).  While the variations do demonstrate the existence
of an uncertainty due to theory, the only correct way to assign a
sensible error
is to compare an NLO calculation (which should already be reasonably
stable with respect to variations of the scale), with a yet-higher order
calculation.  Such estimates will require NNLO calculations.

The other logarithms that arise in perturbative calculations are
infrared ones, associated in perturbation theory with the emission
of soft or nearly-collinear radiation.
In a leading-order calculation,
each jet is modelled by a lone outgoing parton.
An NLO calculation, however, includes contributions with emission
of real radiation.  In these contributions, some jets
 can be made up of two partons.  As jet-defining parameters, such as
the cone size $\Delta R$, are varied, differing fractions of
these contributions will show up as contributions to $n$- or $(n+1)$-jet
cross sections.  This allows
the theoretical prediction to acquire the logarithmic
dependence  on jet-defining parameters exhibited by experimental data.
In addition, the real radiation also gives the leading approximation
to a jet's internal structure.

\sect{Calculations}

\def\pbar{{\overline p}}
\def\L{\left(}\def\R{\right)}
\def\LP{\left.}\def\RV{\right|}
At leading order, each jet is modelled in perturbation theory by a
lone outgoing parton.  A theoretical prediction of an $n$-jet distribution
in hadron-hadron collisions
then requires the probability of finding a parton of given momentum
fraction $x$ inside the nucleon, given by the parton distribution
function $f_{a\leftarrow p}(x,\mu)$, along with
knowledge of the strong coupling $\alpha_s(\mu)$ and the $2\rightarrow n$
tree-level matrix elements.  It also requires a perturbative approximation
to the experimental jet algorithm.  Assembling these ingredients, the
differential distribution in the experimental observable
$X(\{k_j\}_{j\in {\rm jets}})$ is
$$\eqalign{
\LP{d\sigma_n^{\rm LO}\over dX}\RV_{\rm cuts} &=
  \int dx_1 dx_2\;\sum_{ab}
      \int_{\rm cuts} d{\rm LIPS}(x_1 x_2 s;\{k_i\}_{i=1}^n)\,\cr
 &\hskip 20mm\times
   \alpha_s^n(\mu) f_{a\leftarrow p}(x_1,\mu)
     f_{b\leftarrow \pbar}(x_2,\mu)\;
   {\rm JetSelect}(\{k_i\}_{i=1}^n)\cr
&\hskip 30mm\times
{\cal M}(a+b\rightarrow \{f_i\}_{i=1}^n)\, \delta\L X-X(\{k_i\})\R\;,
}\anoneqn$$
where $d{\rm LIPS}$ is the Lorentz-invariant phase space measure,
${\cal M}$ is the tree-level squared perturbative matrix element,
and JetSelect is the perturbative approximation to the experimental
jet algorithm.

In order to study processes with more exclusive final states, such
as processes with specific mesons, we also need a set of
{\it fragmentation functions\/}, which play the opposite role from
distribution functions.  The fragmentation function
$D_{H\leftarrow a}(z,\mu)$ gives the probability of producing a
final-state hadron $H$ with momentum fraction $z$ from the outgoing
parton $a$.  It depends on a renormalization scale $\mu$, which in this
context is called a factorization scale.  Including the fragmentation
function, we would obtain a formula along the lines of
$$\eqalign{
\LP{d^3\sigma_n^{\rm LO}\over dk_H^3}\RV_{\rm cuts} &=
  \int dx_1 dx_2\;\sum_{ab}
   \int_{\rm cuts} d{\rm LIPS}(x_1 x_2 s; \{k_i\}_{i=1}^n)\,\cr
&\hskip 10mm\times
     \alpha_s^n(\mu)\,f_{a\leftarrow p}(x_1,\mu)
     f_{b\leftarrow \pbar}(x_2,\mu)\;
   {\rm JetSelect}(\{k_i\}_{i=1}^n)\cr
&\hskip 20mm\times
{\cal M}(a+b\rightarrow \{f_i\}_{i=1}^n)\,
\; \sum_c \int dz\; D_{H\leftarrow c}(z,\mu)\,\delta^3(k_H-z k_1)\,.
}\anoneqn$$

At next-to-leading order, we must combine real-emission contributions
with virtual contributions.  Each of these contributions is independently
singular.
This means that we have to combine the contributions analytically,
while performing the phase-space integrals numerically.  There are
two basic approaches to this problem.  One, the so-called
`slicing' method, is to separate the
real-emission phase into two regions.  In the soft or collinear region, the
the integral is calculated analytically, using [universal]
soft or collinear approximations.  In the remaining region, the integral
is finite and can be calculated numerically.  The result of integrating
the real-emission contribution over the soft and collinear phase space
can be combined with the virtual contribution; the sum is again finite,
and the integral over [hard] phase space can be performed numerically.
For the general version of this approach, see
refs.~[\ref\GG{W. T. Giele and E. W. N. Glover,
        Phys. Rev. {D46}:1980  (1992)\semi
W. T. Giele, E. W. N. Glover and D. A. Kosower,
        Nucl. Phys. {B403}:633 (1993) [hep-ph/9302225]}].

In the other approach, one subtracts an approximation to the real-emission
matrix element everywhere in phase space.  The approximation is designed
so that the integral
factors into an analytically doable (but singular) integration
times a phase space
integral which can be evaluated numerically.  The singularities again cancel
the singularities in the virtual corrections to the matrix element.  The
integration of the original matrix element less the subtrahend is finite,
and again can be performed numerically.  The reader will find general
versions of this approach in refs.~[\ref\KST{S. Frixione, Z. Kunszt, and A.
Signer, hep-ph/9512328},\ref\CS{S. Catani and M. Seymour, hep-ph/9602277 and
hep-ph/9605323}].

\def\eind{\hskip 3mm}
Within the general slicing method, one can schematically write the
NLO form of a differential cross section as follows,
$$\hskip -5mm\eqalign{
\LP{d\sigma_n^{\rm NLO}\over dX}\RV_{\rm cuts}\hskip -2mm =&
  \int dx_1 dx_2\;\sum_{ab}  \alpha_s^n(\mu)\,\Bigl\{
  f_{a\leftarrow p}(x_1,\mu)
  f_{b\leftarrow \pbar}(x_2,\mu)\,\hat\sigma^{\rm LO}(x_{1,2} \rightarrow n)
   \cr
&\eind + \alpha_s(\mu) f_{a\leftarrow p}(x_1,\mu)
  f_{b\leftarrow \pbar}(x_2,\mu)\,
    K(x_1,x_2)\otimes_{\rm perm}\hat\sigma^{\rm LO}(x_{1,2} \rightarrow n)\cr
&\eind + \alpha_s(\mu) \left[ C_{a\leftarrow p}(x_1,\mu)
  f_{b\leftarrow \pbar}(x_2,\mu)
  +f_{a\leftarrow p}(x_1,\mu)
  C_{b\leftarrow \pbar}(x_2,\mu)\right]\,
    \hat\sigma^{\rm LO}(x_{1,2} \rightarrow n)\cr
&\eind
  +\alpha_s(\mu) f_{a\leftarrow p}(x_1,\mu)
  f_{b\leftarrow \pbar}(x_2,\mu)\,\hat\sigma^{\rm LO}(x_{1,2} \rightarrow (n+1)
  \;{\rm finite})
   \cr
&\eind
  +\alpha_s(\mu) f_{a\leftarrow p}(x_1,\mu)
  f_{b\leftarrow \pbar}(x_2,\mu)\,
   \hat\sigma^{\rm NLO\ finite}(x_{1,2} \rightarrow n)\Bigr\}\,.
   \cr
}\anoneqn$$
In this equation, the second term inside the braces summarizes the
contribution of the integration over soft and final-state collinear regions,
once combined with the corresponding virtual singularities.  The third
term summarizes the contribution of initial-state collinear regions,
again along with corresponding virtual singularities.  It makes use
of {\it crossing functions\/} $C_a$, which are essentially convolutions
of the parton distribution functions with the Altarelli-Parisi splitting
functions.  These functions, along with the $K$ function in the second
term, are independent of the short-distance process, and hence do not
have to be calculated anew for each new process one wants to calculate
numerically.
The fourth term gives the contributions of the finite parts
of the real emission contributions, that those outside the soft and
collinear regions.  The last term gives the contributions of the finite
parts of the virtual corrections to the matrix element; it is the complexity
of calculating these corrections that is at present
the limiting factor in writing NLO programs for new processes.

\sect{Single-Jet Inclusive Distribution}

\def\pbar{{\overline p}}
One of the simplest distributions to consider is the single-jet
inclusive distribution.  As its name suggests, one is studying
$$
p\pbar \rightarrow {\rm jet} + X\,,
$$
binning all jets in transverse energy $E_T$.  Of course, most
events that show up in this distribution are actually two-jet events
(the second jet is needed to balance the $E_T$).
Now, production of a pair of jets at high $E_T$ requires a large
partonic center-of-mass energy, and so the high-$E_T$ tail of this
distribution probes the large-$\hat s$ region, where one is most
likely to see signals of new physics.

The CDF collaboration has claimed to find evidence of a
discrepancy\mref\CDFjet{W. Goshaw, CDF collaboration, these proceedings\semi
CDF Collaboration, preprint Fermilab--PUB--96--020--E [hep-ex/9601008]} between
their results and an
NLO calculation\mref\EKS{S. D. Ellis, Z. Kunszt and  D. E. Soper,
Phys.\ Rev.\ Lett.\ 69:3615 (1992)}.  The nature of the
discrepancy depends on which parton distribution set is used in the
theoretical calculation.  CDF chose to use an older
set, and then found that data points at transverse energies
of 250~GeV and up are systematically higher than the data.

Were $\alpha_s$ known to much higher accuracy than it is,
were the gluon distribution in the proton known to much higher accuracy
than it is; and were the discrepancy a remarkable rise at large-$E_T$,
well outside of statistical and systematic errors, one might then
lean towards seeing in it a signal of physics beyond the standard
model, perhaps indeed of compositeness.  Unfortunately, neither $\alpha_s$
nor the gluon distribution are known well enough to draw such conclusions,
and the other stated hypotheses deserve closer scrutiny as well.

The D\O\ collaboration's results, as presented
at this conference by G.~Blazey\mref\Dzero{G. Blazey, D\O\ collaboration, these
proceedings}, fail to confirm the CDF claim.  They don't necessarily
contradict it, either; a more careful comparison of the two data sets
and a more thorough examination of the systematic errors of the D\O\ data
set would be required to draw such a conclusion.

This differential cross section spans an
enormous dynamic range, from $10^4$~nb/GeV at $E_T\sim 60$~GeV
to $10^{-2}$~nb/GeV at $E_T\sim 400$~GeV.  We must bear in mind
that when we view the experimental results in the
form $({\rm data}-{\rm theory})/{\rm theory}$,
certain systematic errors can induce rather large effects.
In particular, the experimental data must be corrected for the
detector's resolution: a real-world detector may report an energy
deposit different from the actual energy of a jet.
Correcting for this resolution
requires shifting jets along the distribution from
one $E_T$ to another; the rapidly-falling distribution magnifies the
results of uncertainties in estimating the tails of the resolution
function.

Aside from possible experimental systematic errors, the
most plausible explanation for the discrepancy is our lack of
sufficiently detailed knowledge of the gluon distribution function
in the nucleon.  As S.~Kuhlmann showed in his
talk\mref\Kuhlmann{S. Kuhlmann, these proceedings},
the use of a different gluon distribution, along with a slightly different
$\alpha_s(M_Z)$, will bring the QCD prediction into agreement with
the CDF results.  (The modified distribution and $\alpha_s(M_Z)$ still
agree with deeply inelastic scattering data.)

Other theoretical ``explanations'' of the excess seem to me much less
plausible. Even at the highest energies in the CDF distribution, $x_T$
is at most of order $0.5$.  These points are thus far from the kinematic
endpoint, and resummation of end-point logarithms (of the generic form
$\alpha_s \ln (1-x_T)$) seem unlikely to contribute a 50\% effect.
Higher-order corrections also seem an unlikely candidate; while the
NLO-to-LO ratio depends sensitively on the way the renormalization scale
is chosen, with a natural scale choice of $\mu\sim {\cal O}(E_T)$, this
ratio is flat over a wide $E_T$ range, and thus cannot explain the
change of {\it shape\/} the CDF data seemingly require.

It will, of course, be interesting to see other distributions --- such as
the dijet angular distribution --- which would generically differ substantially
from QCD predictions at high $E_T$ were new physics to show up.

\sect{Quarkonia}

Charmonium, and to a lesser extent bottomonium, production at
hadron colliders are potentially useful probes.  Charmonium production
also plays an important role in collider studies of $B$ physics, which
are in turn promising for studies of $CP$ violation.

\def\cbar{{\overline c}}
For many years, theorists assumed that charmonium production was
dominated by perturbative $c\cbar$
production\mref\ccp{R. Baier and R. R\"uckl, Z. Phys.\ C19:251 (1983)}.
CDF data disagreed wildly with these expectations.
Braaten and Yuan showed, however, that at large transverse momentum,
fragmentation is actually the dominant production
process\mref\BY{E. Braaten and Yuan, \PRL 71:1673 (1993)}.
Recent CDF data, presented by
V.~Papadimitriou\mref\charmonium{V. Papadimitriou, CDF collaboration, these
proceedings}, now distinguish between direct $J/\Psi$ production, and
production
from $B$ decay.  These data still show a much larger rate
than would be predicted from the fragmentation contribution
assuming the latter is dominated by production of color-singlet
charmonium states.  Including color-octet production\mref\octet{
K. Sridhar, these proceedings\semi
P. Cho and A. K. Leibovich, Phys.\ Rev.\ D53:6203 (1996) [hep-ph/9511315]}
 seems to bring the theory into much better agreement with the data
at the Tevatron, though it is not yet clear that we have obtained
a picture consistent with HERA data\mref\HERAcharm{M. Cacciari and M. Kr\"amer,
preprint DESY-96-005 [hep-ph/9601276]}.

\sect{Prompt Photons}

The study of prompt photon production,
$$
p\pbar \rightarrow \gamma + X\,,
$$
is in principle a good way to extract information about the gluon
distribution in the proton\mref\PhotA{%
W. Vogelsang and A. Vogt, Nucl.\ Phys.\ B453:334 (1995) [hep-ph/9505404]}.
In practice, it has been plagued by
problems concerning an appropriate experimental and theoretical
definition of ``photons''.  Because of potential problems with contamination
from $\pi^0 \rightarrow \gamma\gamma$ with overlapping photons inside
the detector, experimenters do not try to observe photons inside
jets.  Instead, they demand that photons be isolated away from jets.

However, from a theoretical point of view, a total isolation cut (no
hadronic energy inside a cone surrounding the candidate photon direction)
is a bad idea, because such a cross section is divergent in perturbation
theory.  (Chopping out a cone in phase space prevents the real-emission
contributions from cancelling all of the singularities in
the virtual corrections.)  As a result, such a cross section is
very sensitive to long-distance, that is non-perturbative, physics.

A theoretically more satisfactory approach is to restrict the
hadronic energy {\it fraction\/} inside the cone, though
recent papers have raised questions about the detailed cancellations
here as well\mref\BGQ{%
E. L. Berger, X. Guo, and J. Qiu, Phys.\ Rev.\ Lett.\ 76:2234 (1996)
[hep-ph/9512281]\semi
E. L. Berger, X. Guo, and J. Qiu, preprint ANL-HEP-PR-96-37 [hep-ph/9605324]}.
 Even with the more
theoretically satisfactory definition, however, the
measurements\mref\PhotonData{A. Maghakian, CDF collaboration, these
proceedings\semi
D. Buchholz, D\O\ collaboration, these proceedings} seem
to lie above the theoretical predictions\mref\PhotPred{%
M. Gluck, L. E. Gordon, E. Reya, and W. Vogelsang,
Phys.\ Rev.\ Lett.\ 73:388 (1994)} at $E_T$ below $30$~GeV.
In the case of the D\O\ data, one may be tempted to ascribe the
disagreement to the larger experimental systematic errors at low $p_T$,
but for the CDF data this doesn't work.
Adding parton showering\mref\Shower{%
H. Baer and M. H. Reno, preprint FSU-HEP-951229, UIOWA-95-07 [hep-ph/9603209]}
 to, or putting in an
intrinsic $k_T$ into the theoretical calculation\cite\Kuhlmann
(both in an ad hoc way) brings the predictions into
better agreement with the data.  While this may provide clues to
a resolution of this discrepancy, it cannot be considered satisfactory
in itself.

\sect{Jet Algorithms}

Measuring jet cross sections requires a precise
definition of a jet.  A jet algorithm, as used by experimenters,
must specify how to cluster the sprays of hadrons observed in the
detector into jets.  It must also have a matching version to be
used by theorists, which specifies how to cluster partons in a
perturbative calculation into jets.

In principle, any infrared-safe algorithm could be used to compare
experimental data with perturbative calculations.  In order to make
the best use of data, however, it is best to choose an algorithm
with good theoretical properties, in particular with small higher-order
corrections.

In $e^+e^-$ annihilation, such considerations have played an important
role in the shift from the traditional JADE or invariant-mass algorithm
to the so-called $k_{\perp}$ or Durham algorithm\mref\Durham{S. Catani,
Yu. L. Dokshitser, M. Olsson, G. Turnock, and B. R. Webber,
Phys.\ Lett.\ B269:432 (1991)}.  The latter allows resummation, and is
expected to have smaller power hadronization corrections and better
mass resolution than the JADE algorithm\mref\Webber{B. R. Webber, in
{\it Proceedings of the 27${}^{\sevenrm th}$ International Conference on
High Energy Physics (ICHEP), Glasgow, Scotland, July 20-27, 1994\/};
[hep-ph/9410268]}.

In hadron-hadron collisions, in contrast, both collaboration use variants
of the so-called `Snowmass' cone algorithm\mref\Snowmass{%
S. D. Ellis, J. Huth, N. Wainer,
K. Meier, N. Hadley, D. Soper, and M. Greco, in {\it Research Directions
for the Decade\/}, Proceedings of the Summer Study, Snowmass,
Colorado, 1990, ed.\ E. L. Berger (World Scientific, Singapore, 1992)}.
  While this algorithm presumably has better properties than the JADE
algorithm, there are aspects of it that are poorly modelled in low orders
of perturbation theory.  In particular, in the experimental algorithm
one is sometimes faced with the choice of `splitting' a jet which contains
two distinct centers.  This cannot be modelled in an NLO calculation (the
simplest perturbative approximation requires three partons forming the
proto-jet, hence an NNLO calculation), and is thus a source of uncertainty
in the theoretical calculation.  The hadronic version of the $k_T$
algorithm\mref\HadronicKT{S. Catani, Yu. L. Dokshitser, B. R. Webber,
Phys.\ Lett.\ B285:291 (1992)\semi
S. Catani, Yu.L. Dokshitser, M. H. Seymour, B. R. Webber,
Nucl.\ Phys.\ B406:187 (1993)}
 avoids this problem, because the $\eta$--$\phi$
plane is not split up into rigid circles as in the cone algorithm, but
rather into odd-shaped regions that adapt to the shapes of the
jets in a given event.  This algorithm presumably shares many of
the features of its $e^+e^-$ forebear, such as better mass
resolution\mref\HadronAlgs{M. H. Seymour, in {\it Proceedings of the
 10${}^{\tenrm th}$ Topical Workshop on Proton-Antiproton Collider
Physics, Batavia, IL, May 9-13, 1995\/}; [hep-ph/9506421]\semi
J. Pumplin,  preprint MSU-HEP-60605 [hep-ph/9606236]}.

\listrefs

\bye